\documentclass{elsarticle}

\usepackage{lineno}

\usepackage{graphicx}
\usepackage{amssymb}

\renewcommand{\theta}{\vartheta}
\renewcommand{\epsilon}{\varepsilon}

\usepackage{color}

\begin{document}

%\tableofcontents
%\newpage

%\linenumbers

\begin{frontmatter}

\title{Measurement of the atmospheric muon depth intensity relation with the NEMO Phase-2 tower}

 \author[INFNCT]{S. Aiello}
 \author[INFNRM]{F. Ameli}
 \author[INFNGE]{M. Anghinolfi}
 \author[INFNNA,UniNA]{G. Barbarino}
 \author[INFNBA,UniBA]{E. Barbarito}
 \author[INFNNA,UniNA]{F. Barbato}
 \author[INFNPI,UniPI]{N. Beverini}
 \author[INFNLNS]{S. Biagi}
 \author[INFNPI]{B. Bouhadef}
 \author[INFNSA,UniSA]{C. Bozza}
 \author[INFNLNS]{G. Cacopardo}
 \author[INFNPI,UniPI]{M. Calamai}
 \author[INFNLNS]{C. Cal\`{\i}}
 \author[INFNRM,UniRM]{A. Capone}
 \author[INFNLNS]{F. Caruso}
 \author[INFNBA]{A. Ceres}
 \author[INFNBO]{T. Chiarusi}
 \author[INFNBA]{M. Circella}
 \author[INFNLNS]{R. Cocimano}
 \author[INFNLNS]{R. Coniglione}
 \author[INFNLNS]{M. Costa}
 \author[INFNLNS]{G. Cuttone}
 \author[INFNLNS]{C. D'Amato}
% \author[INFNLNS]{V. D'Amato}
 \author[INFNLNS]{A. D'Amico\fnref{Nikhef}}
 \author[INFNRM]{G. De Bonis}
 \author[INFNLNS]{V. De Luca}
 \author[INFNNA]{N. Deniskina}
 \author[INFNNA,UniNA]{G. De Rosa}
 
 \author[UniNA]{F. Di Capua} 
 
\author[INFNLNS]{C. Distefano\corref{corr}}
\cortext[corr]{Corresponding author}
\ead{distefano\_c@lns.infn.it}

 \author[INFNRM,UniRM]{P. Fermani}
 \author[INFNPI]{V. Flaminio}
 \author[INFNBO,UniBO]{L.A. Fusco}
 \author[INFNNA,UniNA]{F. Garufi}
 \author[INFNCT]{V. Giordano}
% \author[INFNLNS]{G. Giovanetti}
 \author[INFNLNS]{A. Gmerk} 
 \author[INFNLNS]{R. Grasso} 
 \author[INFNSA,UniSA]{G. Grella}
 \author[INFNGE]{C. Hugon}
 \author[INFNLNS]{M. Imbesi}
 \author[INFNLNS]{V. Kulikovskiy}
 \author[INFNLNS]{G. Larosa}
 \author[INFNLNS]{D. Lattuada}
 \author[INFNLNS]{K.P. Leismueller}
 \author[INFNCT]{E. Leonora}
 \author[INFNLNS]{P. Litrico}
 \author[INFNRM]{A. Lonardo}
 \author[INFNCT]{F. Longhitano}
 \author[INFNCT,UniCT]{D. Lo Presti}
 \author[INFNPI,UniPI]{E. Maccioni}
 \author[INFNBO,UniBO]{A. Margiotta}
 \author[INFNLNF]{A. Martini}
 \author[INFNRM,UniRM]{R. Masullo}
 \author[INFNNA]{P. Migliozzi}
 \author[INFNLNS]{E. Migneco}
 \author[INFNLNS]{A. Miraglia}
 \author[INFNNA]{C.M. Mollo}
 \author[INFNBA]{M. Mongelli}
 \author[INFNPI,UniPI,NAVAL]{M. Morganti}
 \author[INFNGE]{P. Musico}
 \author[INFNLNS]{M. Musumeci}
 \author[INFNRM]{C.A. Nicolau}
 \author[INFNLNS]{A. Orlando}
 \author[INFNLNS]{R. Papaleo}
 \author[INFNBO,UniBO]{C. Pellegrino}
 \author[INFNLNS]{M.G. Pellegriti}
 \author[INFNRM,UniRM]{C. Perrina}
 \author[INFNLNS]{P. Piattelli}
 \author[INFNCT,UniCT]{C. Pugliatti}
 \author[INFNLNS]{S. Pulvirenti}
 \author[INFNGE]{A. Orselli}
 \author[INFNPI,UniPI]{F. Raffaelli}
 \author[INFNCT]{N. Randazzo}
 \author[INFNLNS]{G. Riccobene}
 \author[INFNLNS]{A. Rovelli}
 \author[INFNGE,UniGE]{M. Sanguineti}
 \author[INFNLNS]{P. Sapienza}
 \author[INFNLNS]{V. Sciacca} 
 \author[INFNBA]{I. Sgura}
 \author[INFNRM]{F. Simeone}
 \author[INFNCT]{V. Sipala}
 \author[INFNLNS]{F. Speziale}
 \author[INFNLNS]{M. Spina}
 \author[INFNLNS]{A. Spitaleri}
 \author[INFNBO,UniBO]{M. Spurio}
 \author[INFNSA,UniSA]{S.M. Stellacci}  
 \author[INFNGE,UniGE]{M. Taiuti}
 \author[INFNPI,UniPI]{G. Terreni}
 \author[INFNLNF]{L. Trasatti}
 \author[INFNLNS]{A. Trovato}
 \author[INFNCT]{C. Ventura}
 \author[INFNRM]{P. Vicini}
 \author[INFNLNS]{S. Viola}
 \author[INFNNA,UniNA]{D. Vivolo}

\address[INFNBA]{INFN Sezione Bari, Via E. Orabona 4, 70126, Bari, Italy}
\address[INFNBO]{INFN Sezione Bologna, V.le Berti Pichat 6/2, 40127, Bologna, Italy}
\address[INFNLNS]{INFN Laboratori Nazionali del Sud, Via S.Sofia 62, 95123, Catania, Italy}
\address[INFNCT]{INFN Sezione Catania, Via S. Sofia 64, 95123, Catania, Italy}
\address[INFNLNF]{INFN Laboratori Nazionali di Frascati, Via Enrico Fermi 40, 00044, Frascati (RM), Italy}
\address[INFNGE]{INFN Sezione Genova, Via Dodecaneso 33, 16146, Genova, Italy}
\address[INFNNA]{INFN Sezione Napoli, Via Cintia, 80126, Napoli, Italy}
\address[INFNPI]{INFN Sezione Pisa, Polo Fibonacci, Largo Bruno Pontecorvo 3, 56127, Pisa, Italy}
\address[INFNRM]{INFN Sezione Roma, P.le A. Moro 2, 00185, Roma, Italy}
\address[INFNSA]{INFN Gruppo Collegato di Salerno, Via Giovanni Paolo II 132, 84084 Fisciano, Italy}

\address[UniBA]{Dipartimento Interateneo di Fisica Universit\`a di Bari, Via E. Orabona 4, 70126, Bari, Italy}
\address[UniBO]{Dipartimento di Fisica Universit\`a di Bologna, V.le Berti Pichat 6/2, 40127, Bologna, Italy}
\address[UniCT]{Dipartimento di Fisica e Astronomia Universit\`a di Catania, Via S. Sofia 64, 95123, Catania, Italy}
\address[UniGE]{Dipartimento di Fisica Universit\`a di Genova, Via Dodecaneso 33, 16146, Genova, Italy}
\address[UniNA]{Dipartimento di Scienze Fisiche Universit\`a di Napoli, Via Cintia, 80126, Napoli, Italy}
\address[UniPI]{Dipartimento di Fisica Universit\`a di Pisa, Polo Fibonacci, Largo Bruno Pontecorvo 3, 56127, Pisa, Italy}
\address[UniRM]{Dipartimento di Fisica Universit\`a ``Sapienza'', P.le A. Moro 2, 00185, Roma, Italy}
\address[UniSA]{Dipartimento di Fisica Universit\`a di Salerno}
\address[NAVAL]{Accademia Navale di Livorno, viale Italia 72, 57100 Livorno, Italy}

\fntext[Nikhef]{Present address: Nikhef, Science Park, Amsterdam, The Netherlands}

\begin{abstract}

The results of the analysis of the data collected with the NEMO Phase-2 tower, deployed at 3500 m depth about 80 km off-shore Capo Passero (Italy), are presented. \v{C}erenkov photons detected with the photomultipliers tubes were used to reconstruct the tracks of atmospheric muons. Their zenith-angle distribution was measured and the results compared with Monte Carlo simulations.  An evaluation of the systematic effects due to uncertainties on environmental and detector parameters is also included. The  associated depth intensity relation was evaluated and compared with previous measurements and theoretical predictions. With the present analysis, the muon depth intensity relation has been measured up to 13 km of water equivalent.

\end{abstract}

\end{frontmatter}

\section{Introduction}
\label{sec:introduction}

High energy neutrinos are considered optimal probes to identify the sources of high energy cosmic rays. Therefore the search for neutrino sources is one of the most interesting topics of astroparticle physics.  Given the low expected neutrino fluxes from galactic and extragalactic sources \cite{modelli}, high energy neutrino astronomy requires km$^3$-scale detectors. After the first generation of neutrino telescopes, such as BAIKAL \cite{baikal}, AMANDA \cite{amanda} and ANTARES \cite{antares}, the IceCube telescope, at the South Pole, recently showed the first evidence for high energy extraterrestrial  neutrinos \cite{icecube}. 

The next step will be the construction in the Mediterranean Sea of a deep-sea km$^3$-scale  detector: the KM3NeT telescope \cite{km3net}. The northern hemisphere is optimal to observe the Galactic region through the $\nu_\mu$ charged current interaction channel, which guarantees sub-degree resolution on the determination of the source position.
KM3NeT is the result of a joined effort of the ANTARES, NEMO and NESTOR Collaborations, which conducted intense R\&D activities.
KM3NeT will be implemented as a distributed infrastructure in three sites: Toulon (France) \cite{tolone}, Capo Passero (Italy) \cite{sito} and Pylos (Greece) \cite{pylos}. It has been shown that this solution does not deteriorate the telescope performance in the $\nu_\mu$ channel.

A first small-scale prototype, a tower-like detection unit (NEMO Phase-1 tower), was operated for five months during 2007 at 2080 m depth offshore Catania (Italy) \cite{Phase1}.
In March 2013, a larger scale prototype (NEMO Phase-2 tower) was deployed in the Capo Passero site at a depth of  3458 m. This detector was continuously operated for more than one year.

In this paper, after a description of the detector, we report on the measurement of the atmospheric muon angular distribution and on the comparison with Monte Carlo simulations. The corresponding muon depth intensity relation is evaluated and compared with theoretical predictions and with previous results.
The present analysis extends the muon depth intensity relation measured in water up to 13 km.

\section{The NEMO Phase-2 tower}
\label{sec:tower}

The  NEMO Phase-2 tower was deployed on March 23 2013 at the site located about 80 km offshore Capo Passero (latitude: 36$^\circ$ 17' 48" N, longitude: 15$^\circ$ 58' 57'' E) at a depth of 3458 m. The tower operated continuously until August 4 2014, when it was disconnected to allow for an upgrade of the infrastructure.

\begin{figure}[h]
\begin{center}
\includegraphics[width=8cm]{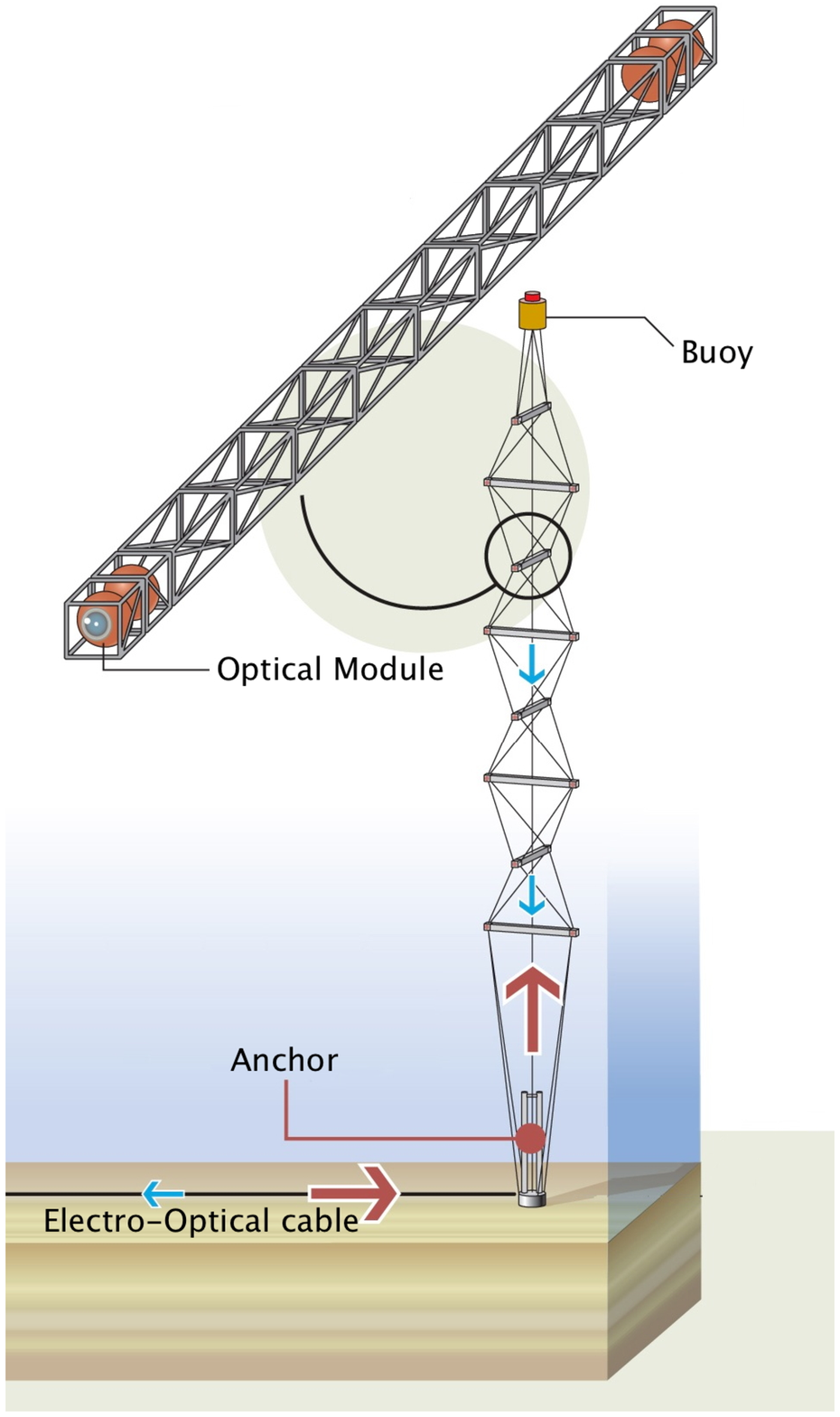}
\end{center}
\caption{Artistic view of the NEMO Phase-2 tower. Vertical distances are not to scale.}
\label{fig:tower}
\end{figure}

The tower is a three-dimensional flexible structure composed of eight horizontal elements ({\it floors}) interlinked by a system of tensioning ropes and anchored at the seabed. The structure is kept taut by a system of buoys at the top.
The tower floor, a 8 m long structure, is connected to the next ones with four ropes such that each floor is perpendicular to its vertical neighbours as shown in Fig. \ref{fig:tower}. 
The floors are vertically spaced by 40 m, with the lowermost one located 100 m above the sea bottom. Each floor supports four Optical Modules (OMs), two at each end, one down-looking and one horizontally looking. The three-dimensional structure of the tower provides an unambiguous reconstruction of the muon direction even with a single detection unit.

An OM is composed by a photo-multiplier tube (PMT) enclosed in a 13" pressure resistant glass sphere. The PMT is a 10"  Hamamatsu\footnote{Hamamatsu Photonics, 812 joko-cho, Hamamatsu city, 431-31
Japan, web-site: www.hamamatsu.com} R7081Sel with 10 dynode stages. 
In spite of its large photocathode area, this model has a good time resolution, with a transit time spread of about 3 ns as FWHM for single photo-electron (s.p.e.) pulses, a 35\% charge resolution as sigma and about 25\% quantum efficiency at 400 nm wavelength \cite{pmt}.
Mechanical and optical contact between the PMT and the internal glass surface is ensured by an optical silicone gel. A $\mu$-metal cage shields the PMT from the Earth magnetic field.
The high voltage distribution board (ISEG\footnote{ISEG Bautzer Landstr. 23, 01454 Radeberg / OT Rossendorf, Germany} PHQ 7081SEL) requires a low voltage supply (+5 V) and generates all voltages for cathode, grid and dynodes with a power consumption lower than 150 mW \cite{Nuccio}.

A Front-End Module board (FEM) \cite{Nicolau06} is also placed inside the OM. It applies a quasi-logarithmic compression on the analog signal, which is then sampled at 200 MHz by means of two 100 MHz Flash ADCs, staggered by 5 ns.
A  Field Programmable Gate Array (FPGA) classifies valid samples, stores them with an event time stamp in an internal 16 kb FIFO, packs all OM data and local slow control information and codes everything into a bit stream frame transmitted on a differential pair at 20 Mb/s. 
The board digitizes and transfers pulse waveform information up to a maximum continuous rate of  $\sim$150 kHz.
Moreover, the board has embedded electronics, analog and digital, to control the OM power supply, to monitor temperature, relative humidity and electrical parameters.

Phase-2 tower hosts also several sensors for calibration and environmental monitoring: a
Doppler Current Sensor to measure water current; a light transmissometer to measure water transparency; two Conductivity--Temperature--Depth probes to monitor sea water properties; a pair of hydrophones on each floor and on the tower base for acoustic positioning \cite{MariaGrazia}. Slow control data (including data from environmental sensors and the acoustic positioning system) are checked from shore by means of a dedicated Slow Control Management System \cite{Rovelli06}. 

\section{The on-line muon trigger}

The presence of an optical background, due to $^{40}$K decays and bioluminescence, gives an average rate of about 50 kHz of uncorrelated s.p.e. signals on each PMT \cite{MariaGrazia}. An on-line selection algorithm is used to reduce the amount of data due to the background. This algorithm is based on searching {\em trigger seeds} among the PMT signals ({\em hits}). The trigger seeds consist of time coincidences between PMTs or high charge hits.  
Three different Level 1 (L1) trigger seeds were used:

\begin{itemize}
\item Simple Coincidence (SC): a coincidence, within 20 ns, between two hits occurring on two PMTs located at the same extremity of a given floor;
\item Floor Coincidence (FC): a coincidence, within 100 ns, between two hits recorded at the opposite ends of the same floor;
\item Charge Shooting (CS): a hit exceeding a charge threshold of about 10 p.e.
\end{itemize}

When a L1 trigger seed is found, the data acquired by all the PMTs within the {\em Triggered Time Window} (TTW) are selected, defining an {\em event}. 
The length of the  TTW is fixed at $\pm$3 $\mu$s around the trigger seed time. If a further trigger seed occurs in the TTW after the  first one, the TTW  itself is extended by an additional time interval of 3 $\mu$s after the new seed time.
The event is then stored if one of the following Level 2 (L2) seeds is found:

\begin{itemize}
\item  at least two SCs; 
\item  one SC and one CS in a PMT not participating in the SC;
\item  one FC where one of the two hits participates also in a SC.
\end{itemize}

The measured total on-line trigger rate is about 100 Hz  \cite{Chiarusi14}. From Monte Carlo simulations, the expected rate of atmospheric muons is $\sim0.1$ Hz.

\section{Atmospheric muon data analysis}

\subsection{PMT data calibration}

\begin{figure}
\begin{center}
\includegraphics[width=9cm]{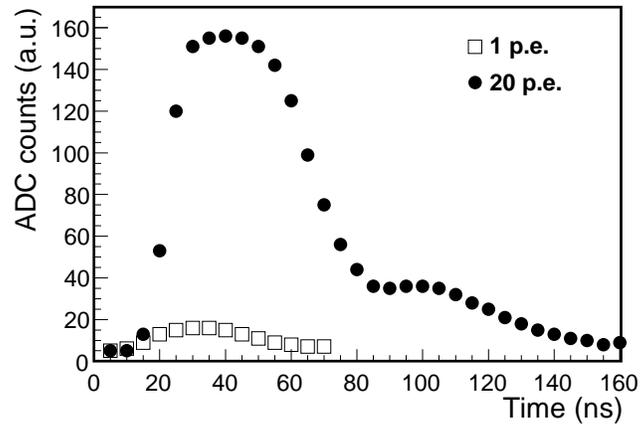}
\includegraphics[width=9cm]{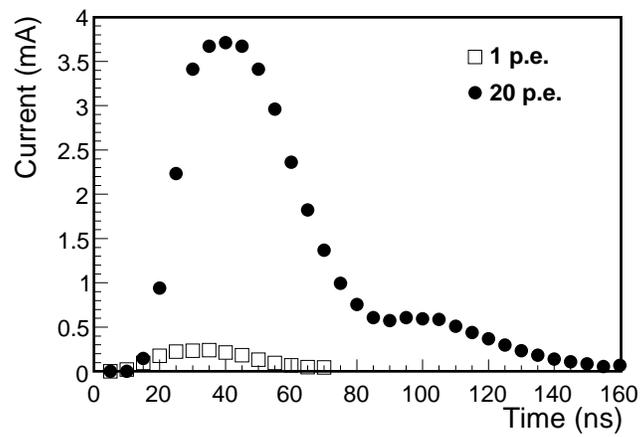}
\end{center}
\caption{Compressed (top panel) and decompressed (bottom panel) waveforms of two particular hits recorded by the tower PMTs.}
\label{fig:wave}
\end{figure}

The first step of the off-line analysis consists of decompression and calibration of PMT hits to convert ADC channels into amplitudes (in mA units) \cite{Ameli08}.
The waveforms of two hits (one s.p.e and one 20 p.e.) before and after the decompression procedure are shown in Fig. \ref{fig:wave}.  

\begin{figure}[h]
\begin{center}
%\vspace{0.8cm}
\includegraphics[width=9cm]{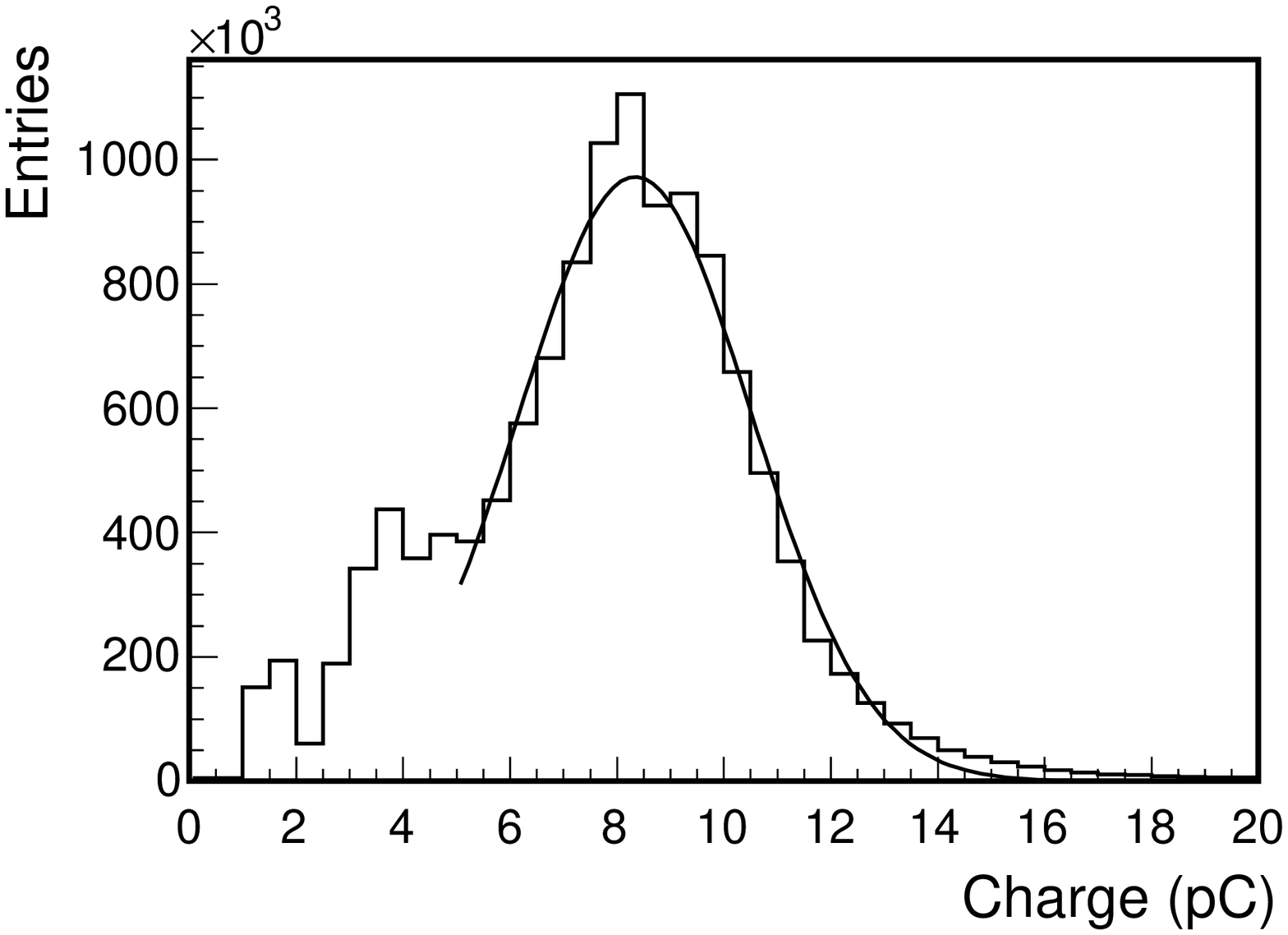}
\end{center}
\caption{Typical charge spectrum of one of the tower PMTs. The spectrum is fitted with a Gaussian function with mean 8.3 pC and width of 2.2 pC.}
\label{fig:pctope}
\end{figure}

PMT high voltages were set in order to have a conversion factor of $\sim$ 8 pC/s.p.e. (see Fig. \ref{fig:pctope}). This value optimises the front-end electronics performance \cite{Ameli08}. 
The total hit charge (in pC) is then computed and converted in units of p.e., using conversion factors determined from the fit of s.p.e. charge spectra. 
During the tower operation time, the PMT gains varied by $\sim10\%$ because of the ageing \cite{Nuccio}. For this reason, the conversion factors are computed on a run-by-run basis during the whole period.

The threshold crossing time (hit time) is evaluated by fitting the waveform leading edge with a sigmoid function. Hit times are corrected removing time offsets measured onshore with a calibration system before to the detector deployment. After the calibration procedure, the accuracy on the hit time is better than 1 ns.

\subsection{The off-line muon filter}
\label{sec:filter}

For each event, all hits participating in simple (SCs) and floor (FCs) coincidences or exceeding a charge of 2.5 p.e. are analysed. In particular, for each hit the number of correlated hits satisfying the condition
\begin{equation}
|dt|<dr/v + 20 \hbox{ ns}, \label{eq:causal}
\end{equation}
is determined. Here, $|dt|$ is the absolute value of the time difference between the hits, $dr$ is the distance between the PMTs where the hits were detected, $v$ is the group velocity of light in seawater.
The maximum number of causality relations $N_{Caus}$ found in each event is used to reject the background. Events with $N_{Caus}\ge6$ are selected for the analysis. Monte Carlo simulations show that this condition increases the muon purity  by three orders of magnitude.

\subsection{Muon track reconstruction}

The events selected by the off-line filter are processed with the ANTARES track reconstruction code \cite{Heijboer03}, adapted to the NEMO tower configuration \cite{Distefano07}. Background hits are rejected applying a causality criterion. The track fitting strategy is based on the maximisation of a likelihood function. The reduced log-likelihood value is denoted as $\Lambda$. The reconstruction algorithm takes into account the \v{C}erenkov light features and the possible presence of unrejected background hits. It starts with a linear pre-fit using all selected hits participating in SCs or exceeding 2.5 p.e. At least three of these hits are required to compute the pre-fit. Starting from the result of the pre-fit, a sequence of fit procedures using all the hits that passed the causality criterion is applied. The nominal PMT positions are assumed; the uncertainties due to this assumption have been evaluated to be negligible for this analysis \cite{SALVO}.

\subsection{Monte Carlo simulation}
\label{sec:simul}

A Monte Carlo simulation of atmospheric muon events was performed using the MUPAGE code \cite{Carminati08}. Bundles of atmospheric muons were generated on the surface of a can-shaped volume of water containing the instrumented volume (573 m height, 310 m radius, the bottom surface located at a depth of 3458 m).

The generated muon events were propagated through the active volume of the detector by using the simulation tools developed by the ANTARES Collaboration \cite{Margiotta13}. These codes simulate the emission and propagation of \v{C}erenkov light induced by muons and their secondary products (e.g. showers and $\delta$-rays), then record photo-electron signals on PMTs. Simulations were performed  on a run-by-run basis, taking into account the experimental conditions of each run. The light absorption length as a function of photon wavelength, previously measured at the detector site \cite{sito}, was taken into account. The OM angular acceptance is the result of an accurate GEANT4 \cite{geant4} simulation. Effects due to shadowing from the OM mechanical support system are included among systematic uncertainties (see Tab. \ref{tab:errors}).

Background hits were added to the hits generated by the atmospheric muon bundle. This background was simulated generating uncorrelated s.p.e. hits with a constant rate of 52 kHz for each PMT, corresponding to the average baseline rate measured experimentally \cite{MariaGrazia}. The DAQ electronics and the on-line trigger were then simulated. Monte Carlo triggered events were processed with the same analysis chain used for the detector data analysis.

\subsection{Results}
\label{sec:results}

Data recorded between April 2013 and August 2014  were analysed.
A total of 606546 atmospheric muon tracks were reconstructed during 411.1 days of live time. 
Table \ref{tab:ratevcut_data} summarizes the number of events analysed at different stages of the analysis chain and of the reconstruction procedure.

\begin{table}[h]
\renewcommand{\arraystretch}{1.2}
\begin{center}
\caption{{\bf Results of the data analysis and track reconstruction procedure:} number of events collected during 411.1 days of live time and surviving each step of the analysis applied in sequence (see text). The second column give  the corresponding Monte Carlo (MC) number of events; the third column gives the data over MC event ratio.
} \label{tab:ratevcut_data} ~\\*[0.2cm]
\begin{tabular}{l|c|c|c}
\hline
													&	{\bf  Data}				& {\bf  MC}		& {\bf  Data/MC} 	\\		
\hline
{\bf  On-line Trigger}            								&  $4.0\times 10^9$ 		 	&  -				& -			\\
{\bf Generated}           								  	&  -						& $ 3.6\times 10^9$  	& -		 	\\
{\bf  Muon Filter}           									&  $1.09\times 10^7$       		& $1.22\times 10^6$	& 8.9	 		\\
{\bf  Reconstructed}              								&  $606546$  				& 491504 			& 1.23		\\
{\bf  Selected ($\Lambda>-10$)}              						&  $269787$  				& 266492 			& 1.01		\\
\hline
\end{tabular}
\\*[1.cm]
\end{center}
\end{table}

\begin{figure}
\begin{center}
\includegraphics[width =9cm]{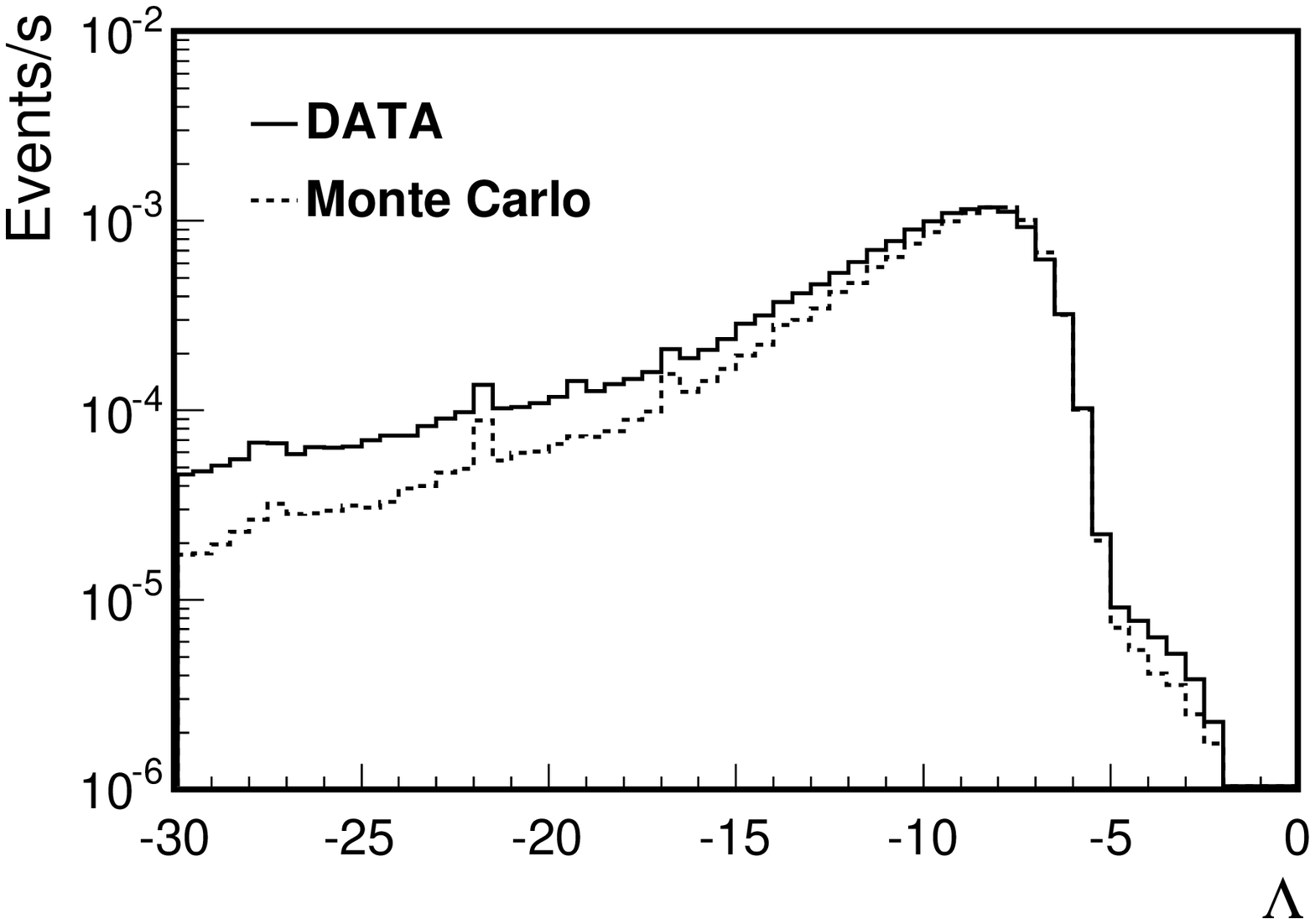}
\end{center}
\caption{Reduced log-likelihood distributions of reconstructed muon tracks for data and Monte Carlo simulation. Events with small values of $\Lambda$ are mainly produced by the background.}
\label{fig:lik}
\end{figure}

\begin{figure}
\begin{center}
\includegraphics[width =9cm]{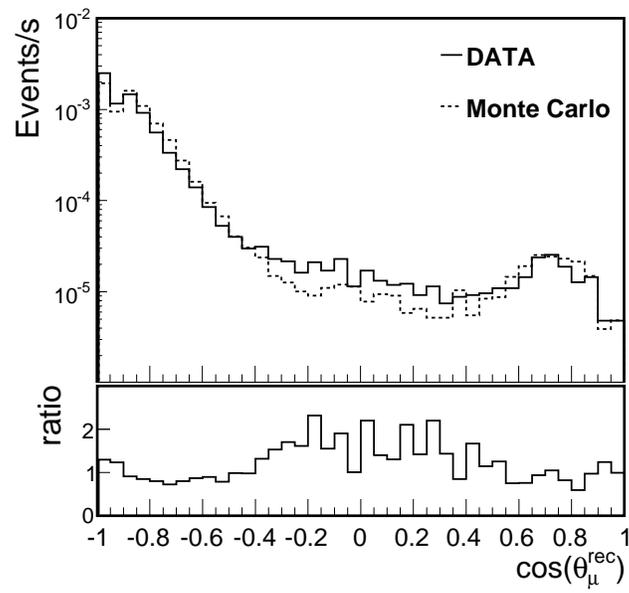}
\end{center}
\caption{Zenith angular distributions of reconstructed atmospheric muon tracks after applying a likelihood quality cut of $\Lambda>-10$. }
\label{fig:muatm}
\end{figure}

\begin{figure}
\begin{center}
\includegraphics[width =9cm]{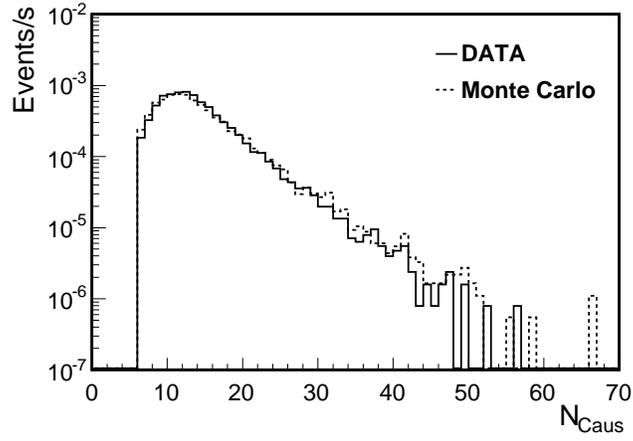}
\end{center}
\caption{Rate of muon tracks, as a function of the off-line muon filter parameter $N_{Caus}$ (see Sec. \ref{sec:filter}), for events surviving the quality cut.}
\label{fig:ncaus}
\end{figure}

\begin{figure}
\begin{center}
\includegraphics[width =9cm]{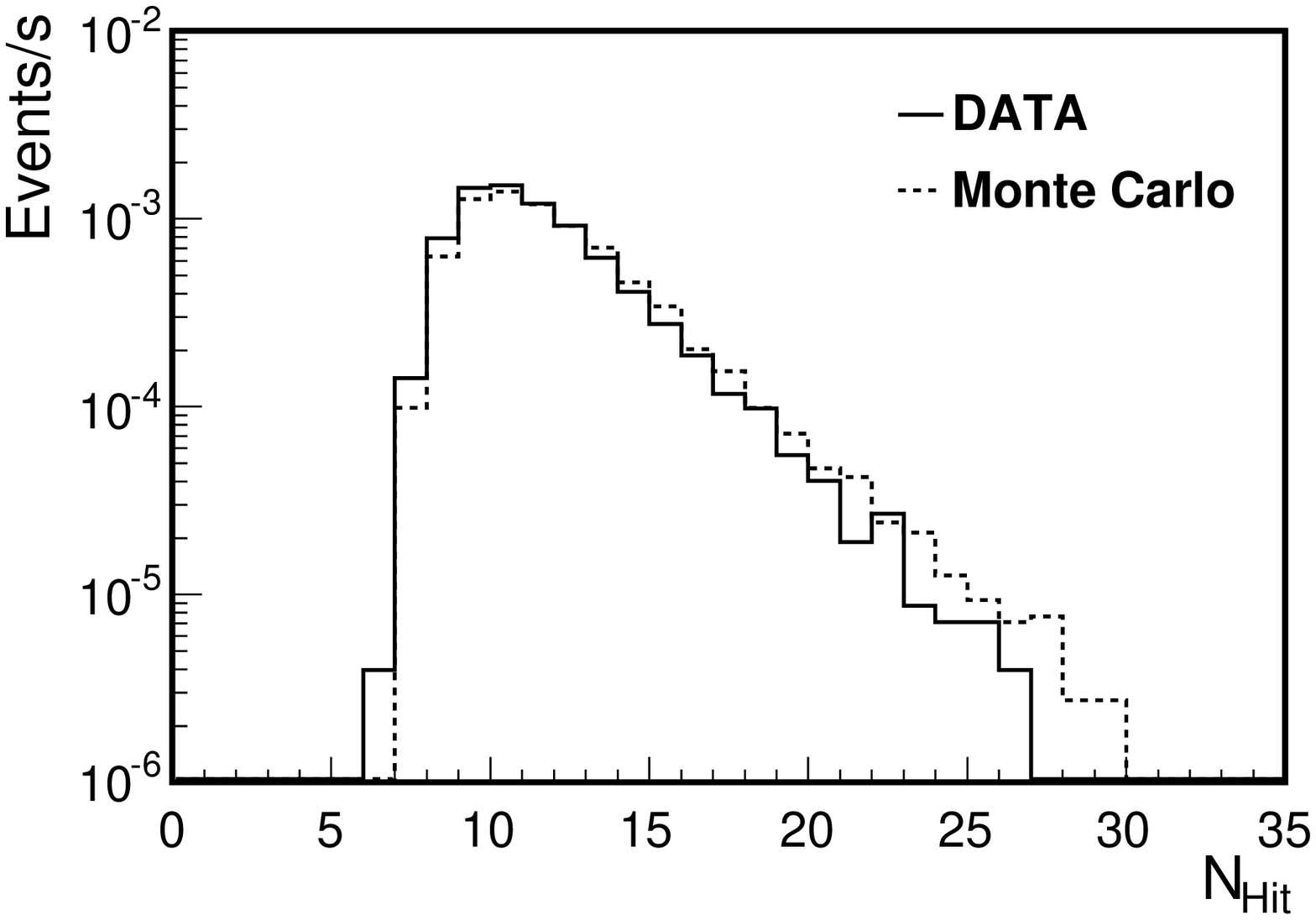}
\end{center}
\caption{Rate of muon tracks, as a function of the number of PMT hits per event used in the reconstruction, for events surviving the quality cut.}
\label{fig:nfit}
\end{figure}

At reconstruction level, the data event rate is 23\% higher than the simulated one. The discrepancy is mainly due to tracks reconstructed with a low likelihood value $\Lambda$ as shown in Fig. \ref{fig:lik};
therefore only the tracks reconstructed with $\Lambda>-10$ were selected. 
After this quality cut a very good agreement is found between data and Monte Carlo simulations, with a data over Monte Carlo event ratio close to one.
The angular distributions of the tracks surviving the quality cut for data and Monte Carlo are shown in Fig. \ref{fig:muatm}. 
A data - Monte Carlo comparison is also shown for the number of hits causally connected by the off-line muon filter $N_{Caus}$  (Fig. \ref{fig:ncaus}) and the number of PMT hits used in the track reconstruction  (Fig. \ref{fig:nfit}).

\section{Depth Intensity Relation for Atmospheric Muons}

The final step of the analysis is the evaluation of the Depth Intensity Relation (DIR) from the reconstructed atmospheric muon tracks. The DIR describes the vertical muon flux intensity as a function of the muon slant depth in water \cite{Phase1}. An atmospheric muon, reaching the detector located at a vertical depth $D$ from a zenith angle $\theta_Z$, propagates through a water slant depth $h$:
\begin{equation}
h=\frac{D}{\cos\theta_Z}.
\end{equation}

The muon intensity $I(\theta_{\mu})$ as a function of the muon track direction $\theta_\mu=180^\circ - \theta_Z$, is calculated using the relation
\begin{equation}
\label{eq:angflux}
I(\theta_{\mu})=  \frac{N_{\mu}(\theta_\mu) \cdot m(\theta_{\mu})} {{A_{eff}}(\theta_{\mu}) \cdot T \cdot d\Omega},
\end{equation}
where

\begin{itemize}

\item  $N_{\mu}(\theta_\mu$) is obtained by applying an unfolding procedure on the reconstructed angular distribution $N_{\mu}(\theta_\mu^{rec})$ and gives the number of events as a function of $\theta_{\mu}$. The iterative unfolding method is based on Bayes' theorem \cite{Dagostini};

\item  $m(\theta_{\mu})$ is the average muon multiplicity as a function of $\theta_{\mu}$, determined from the Monte Carlo simulations as described in section \ref{sec:simul}. $m(\theta_{\mu})$ is shown in Fig. \ref{fig:mult};

\begin{figure}[h]
\begin{center}
\includegraphics[width =9cm]{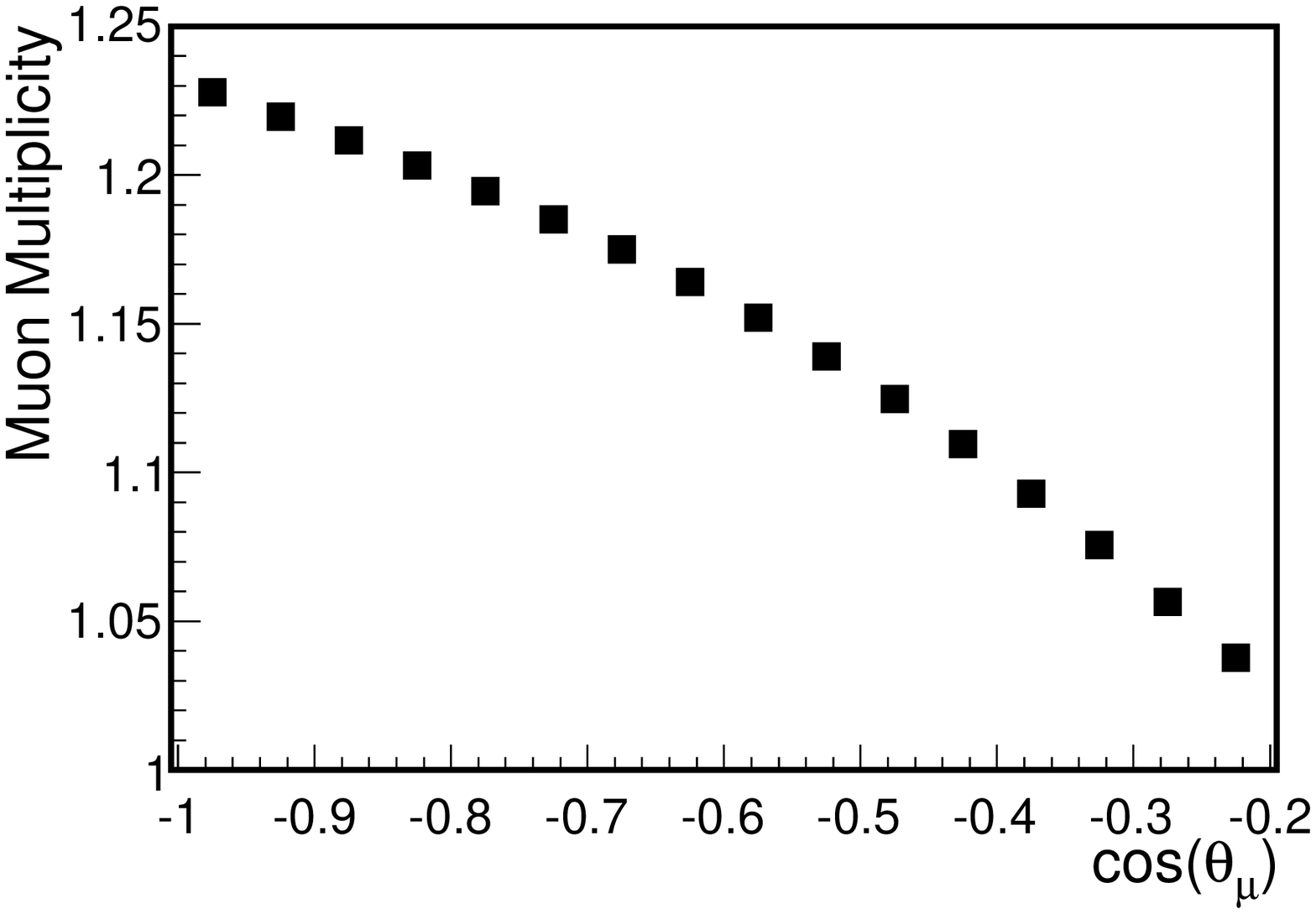}
\end{center}
\caption{Average multiplicity of the muon bundles as a function of $\cos(\theta_{\mu})$ and reaching the detector depth (3458 m) as given by Monte Carlo simulations.}
\label{fig:mult}
\end{figure}

\item $A_{eff}(\theta_{\mu})$ is the detector effective area for reconstructed muon tracks as function of $\theta_{\mu}$ (Fig. \ref{fig:aeff});

\begin{figure}[h]
\begin{center}
\includegraphics[width =9cm]{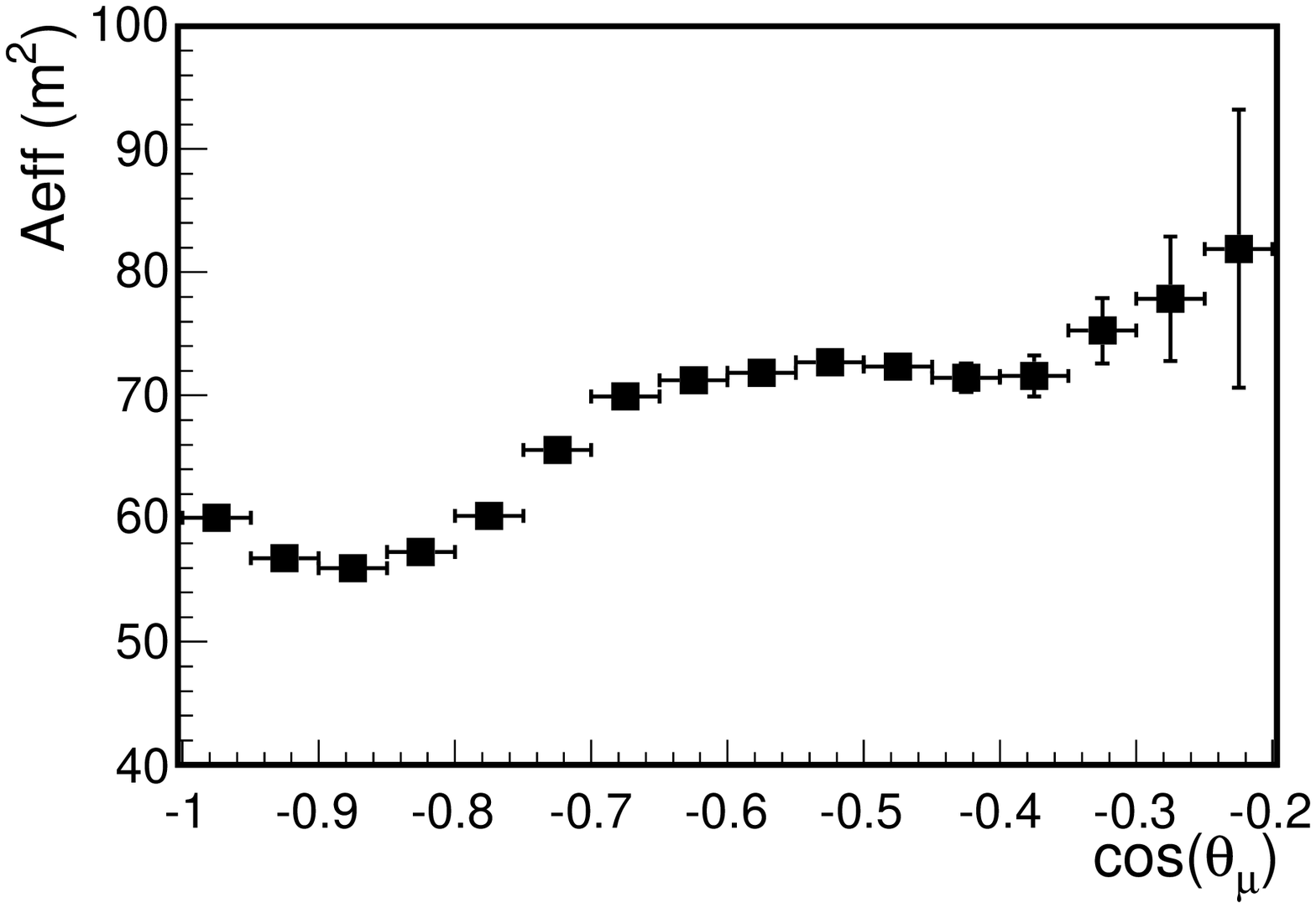}
\end{center}
\caption{Detector effective area for reconstructed muons as a function of $\cos(\theta_{\mu})$ ad determined from Monte Carlo simulations.} \label{fig:aeff}
\end{figure}

\item  $T$ is the detector live time corresponding to the selected data sample; 

\item  $d\Omega$ is the detection solid angle.

\end{itemize}

\begin{figure}[h]
\begin{center}
\includegraphics[width =9cm]{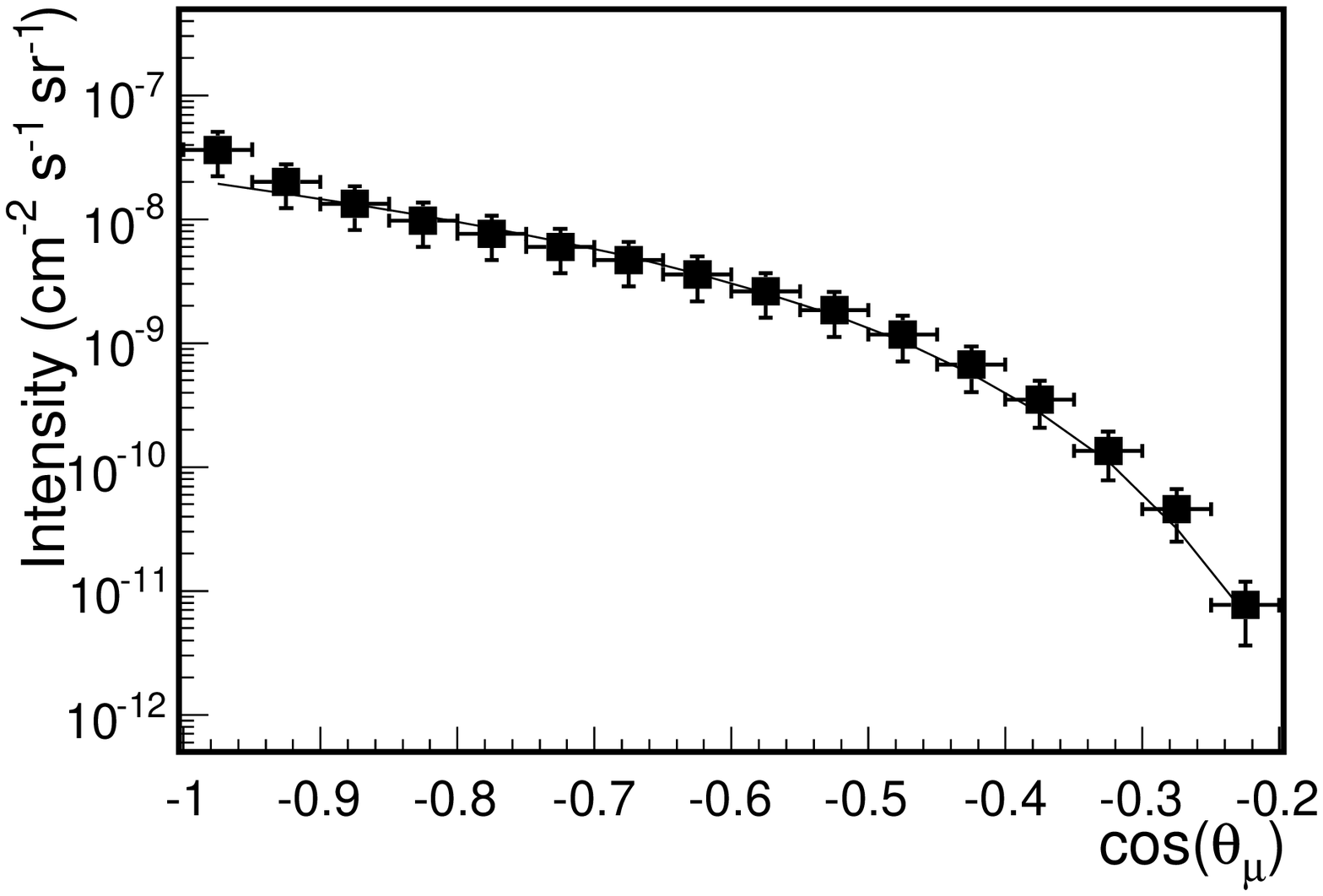}
\end{center}
\caption{Angular distribution of the atmospheric muon flux, $I(\theta_{\mu})$, computed with Eq. \ref{eq:angflux}, for the analysed data period. The error bars include statistical and systematic uncertainties. Data are compared with the simulated atmospheric muon flux (solid line). } \label{fig:angdir}
\end{figure}

The angular distribution of the atmospheric muon flux, obtained from Eq. \ref{eq:angflux}, is shown in Fig. \ref{fig:angdir}. Error bars include statistical and systematic uncertainties. Systematic errors were evaluated via Monte Carlo simulation, taking into account the uncertainties on the input parameters: the light absorption length in water, the light scattering length in water, the PMT quantum efficiency and the OM  angular acceptance \cite{Nuccio}. In Tab. \ref{tab:errors} the contributions of the different parameters to the total systematic error are reported.
The quality of the unfolding procedure is shown by the good agreement with the Monte Carlo simulation of the atmospheric muon flux given by MUPAGE (Fig. \ref{fig:angdir}).

\begin{table}[h]
\renewcommand{\arraystretch}{1.2}
\begin{center}
\caption{{\bf  Systematic errors:}  contribution to the systematic error due to the uncertainty on each input parameter of the Monte Carlo simulation, see text. For OM angular acceptance, systematic errors include the shadowing from tower mechanics as evaluated from GEANT4 simulations (see Sec. \ref{sec:simul}).} \label{tab:errors} ~\\*[0.2cm]
\begin{tabular}{lcc}
\hline
\hline
   {\bf  Input Parameter} & {\bf Relative Uncertainty of the Parameter} & {\bf $\Delta I/I$}  \\
\hline
Light absorption length               & $\pm 10$\%        	&   $^{+15\%}_{-12\%}$  \\
Light scattering length                & $\pm 10$\%        	&   $^{+5\%}_{-2\%}$  \\
PMT quantum efficiency       	& $\pm 10$\%        	&   $^{+20\%}_{-15\%}$  \\
OM angular acceptance        	& -    			&   $^{+30\%}_{-33\%}$   \\
\hline
Total 	&	&	$^{+39\%}_{-38\%}$   \\
\hline
\end{tabular}
\\*[1.cm]
\end{center}
\end{table}

The measured flux $I(\theta_\mu)$ was, then, transformed into muon vertical flux intensity $I(\theta_Z=0,h)$ using:
\begin{equation}
I(\theta_Z=0,h)=I(\theta_Z) \cdot \cos(\theta_Z) \cdot c_{corr}(\theta_Z),
\end{equation}
where the term  $c_{corr}(\theta_Z)$ is a geometrical correction factor which takes into account the curvature of the Earth \cite{Lipari}.

\begin{figure}[h]
\begin{center}
\includegraphics[width =8.5cm]{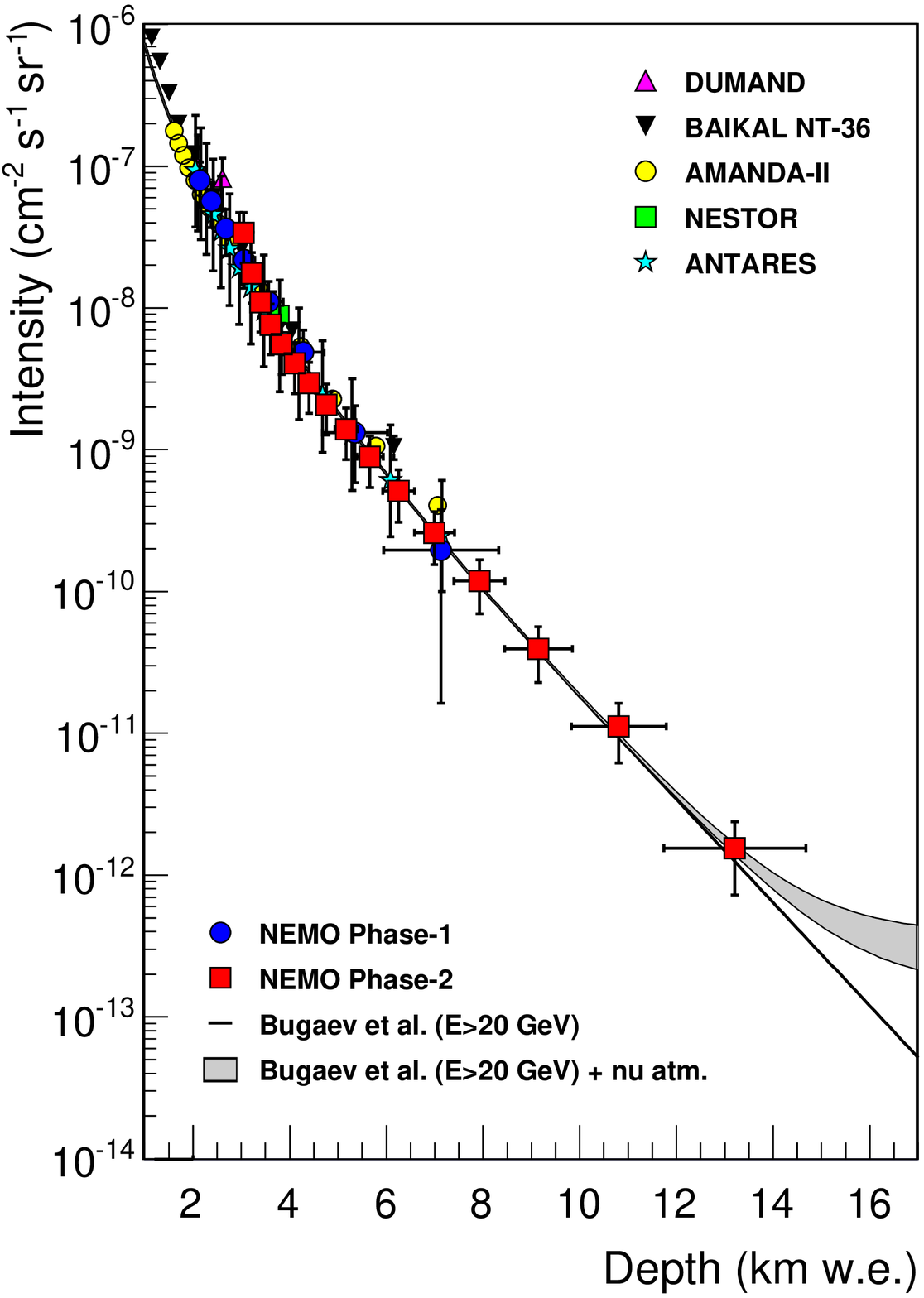}
\end{center}
\caption{Vertical muon intensity, $I(\theta_Z=0,h)$, versus depth measured using data acquired with the NEMO Phase-2 tower. For comparison, results from other experiments are quoted. The solid line is the prediction of  Bugaev et al. \cite{Bugaev}.
The shaded area at large depths includes atmospheric neutrino-induced muons.}
\label{fig:dir}
\end{figure}

The corresponding DIR for atmospheric muons is shown in Fig. \ref{fig:dir}  and tabulated in Tab. \ref{tab:dir}. 
Results obtained by previous experiments are shown for comparison: DUMAND \cite{Dumand}, BAIKAL \cite{Baikal}, NESTOR \cite{Nestor},
 AMANDA \cite{Amanda2}, ANTARES \cite{AntaresDir}  and NEMO Phase-1 \cite{Phase1}. Results are also compared with the
prediction of Bugaev et al. \cite{Bugaev}. NEMO Phase-2 data are well in agreement both with previous measurements and with Bugaev's prediction in the whole range of investigated depths.

\begin{table}[h]
\renewcommand{\arraystretch}{1.2}
\begin{center}
\caption{{\bf  Depth Intensity Relation:} measured vertical muon underwater intensity versus slant depth of water. The quoted errors include statistical and systematic uncertainties. } \label{tab:dir} ~\\*[0.2cm]
\begin{tabular}{ccc}
\hline
\hline
Depth (km w.e.) && Intensity (cm$^{-2}$s$^{-1}$sr$^{-1}$) \\
\hline
$ 3.05\pm 0.08$ &&   $\left(3.4\pm1.3\right)\times 10^{-8} $ \\ 
$ 3.21\pm 0.09$ &&   $\left(1.8\pm0.7\right)\times 10^{-8} $ \\ 
$ 3.40\pm 0.10$ &&   $\left(1.1\pm0.4\right)\times 10^{-8} $ \\ 
$ 3.60\pm 0.11$ &&   $\left(7.6\pm3.0\right)\times 10^{-9} $ \\ 
$ 3.83\pm 0.12$ &&   $\left(5.6\pm2.2\right)\times 10^{-9} $ \\ 
$ 4.10\pm 0.14$ &&   $\left(4.1\pm1.6\right)\times 10^{-9} $ \\ 
$ 4.40\pm 0.16$ &&   $\left(2.9\pm1.2\right)\times 10^{-9} $ \\ 
$ 4.75\pm 0.19$ &&   $\left(2.1\pm0.8\right)\times 10^{-9} $ \\ 
$ 5.17\pm 0.22$ &&   $\left(1.4\pm0.6\right)\times 10^{-9} $ \\ 
$ 5.66\pm 0.27$ &&   $\left(8.9\pm3.5\right)\times 10^{-10} $ \\ 
$ 6.26\pm 0.33$ &&   $\left(5.1\pm2.1\right)\times 10^{-10} $ \\ 
$ 6.99\pm 0.41$ &&   $\left(2.6\pm1.1\right)\times 10^{-10} $ \\ 
$ 7.92\pm 0.53$ &&   $\left(1.2\pm0.5\right)\times 10^{-10} $ \\ 
$ 9.14\pm 0.70$ &&   $\left(3.9\pm1.7\right)\times 10^{-11} $ \\ 
$10.81\pm 0.98$ &&   $\left(1.1\pm0.5\right)\times 10^{-11} $ \\ 
$13.21\pm 1.47$ &&   $\left(1.5\pm0.8\right)\times 10^{-12} $ \\ 
\hline
\end{tabular}
\\*[1.cm]
\end{center}
\end{table}

\section{Conclusions}

The NEMO Collaboration has achieved a major milestone with the installation and operation of a tower-like prototype at 3500 m depth. 
The NEMO Phase-2 tower, composed by 8 floors for a total height of 380 m, equipped with 32 PMTs, was deployed in 2013 about 80 km offshore Capo Passero (Italy).
It was continuously operated for more than one year. Atmospheric muon tracks have been reconstructed and their measured angular distribution has been compared with Monte Carlo simulations.
The muon depth intensity relation has been evaluated and compared with previous data and predictions, showing a good agreement. With the present analysis, the muon
depth intensity relation has been measured in water for the first time up to an equivalent depth of 13 km.

The NEMO Phase-2 tower prototype has validated the technological solutions developed by the collaboration at the operation depth of 3500 m.
After this success, the collaboration started the construction of 8 towers, each with 14 floors, to be installed at the Capo Passero site before the end of 2015. 
These towers will be integral part of the Italian node of the KM3NeT telescope.

\section*{Acknowledgments}

We thank all INFN personnel that has contributed to the development, construction and operation of the mechanics, electronics and computing of the NEMO Phase-2 experiment.

\end{document}